\documentclass[aps]{revtex4}

\def\be {\begin{equation}}
\def\ee {\end{equation}}
\def\ba {\begin{eqnarray}}
\def\ea {\end{eqnarray}}
\def\nn {\nonumber}
\def\a  {\alpha}
\def\b  {\beta}
\def\c  {\gamma}

\def\D  {\Delta}
\def\e  {\epsilon}
\def\la {\label}
\def\le {\left}
\def\ri {\right}

\def\f {\frac}

\def\no {\noindent}
\def\bi {\begin{itemize}}
\def\ei {\end{itemize}}

\def\ra {\rangle}
\def\vs {\vspace}

\begin{document}

\title{Adiabatic Quantum Computation and Deutsch's Algorithm }
\author{ Saurya Das, Randy Kobes, Gabor Kunstatter }
\affiliation{ Physics Dept., The University of Winnipeg and
Winnipeg Institute for Theoretical Physics,
515 Portage Avenue,
Winnipeg, Manitoba - R3B 2E9, CANADA }
\email{saurya,randy,gabor@theory.uwinnipeg.ca  }

\begin{abstract}

We show that by a suitable choice of a time dependent Hamiltonian,
Deutsch's algorithm can be implemented by an adiabatic
quantum computer. We extend our analysis to the Deutsch-Jozsa problem
and estimate the required running time for both global and local
adiabatic evolutions.

\end{abstract}

\vs{.3cm}
\maketitle


Quantum computation and quantum information theory has attracted
a great deal of attention in recent times. Inherently
quantum mechanical systems can in principle be used to implement
a wide variety of computational algorithms wth enhanced
efficiency \cite{review1,review0,review}.
The principle of superposition in quantum mechanics, according
to which a system can be in a linearly superposed state of 
more than one eigenstate, is the key to this increased
efficiency. One of the first algorithms that was first proposed in
this context is Deutsch's algorithm \cite{d}.

In this, one would like to determine whether a function
$$ f : \{0,1\} \rightarrow \{0,1 \} $$ is constant or balanced,
i.e. whether $f(0) = f(1)$ or $f(0) \neq f(1)$ using a
quantum computer.

The four possible outcomes of $f$ are:
\ba
 f(0) = f(1) = 0 ~~~~\mbox{(constant)}  \nn \\
 f(0) = f(1) = 1 ~~~~\mbox{(constant)}  \nn \\
 f(0) = 0 ~,~ f(1) = 1 ~~~~\mbox{(balanced)} \nn \\
 f(0) = 1 ~,~ f(1) = 0 ~~~~\mbox{(balanced)}  \nn
\ea
Ordinarily, one has to determine {\it both} $f(0)$ and $f(1)$
to infer the nature of the function, since the knowledge of one does not
shed light on the value of the other. However, it was shown
that by applying a certain sequence of unitary operators
(`gates') on a given initial {\it quantum mechanical} state, and then
making just {\it one} measurement on the final state, the nature
of the function $f$ can be determined \cite{d}.

Recently, a new framework of quantum computation has been proposed,
in which the series of gates referred to above is entirely replaced by a
Hamiltonian which changes continuously with time. The Hamiltonian
is so chosen that the state of the system is its ground
state at all times (although the ground state itself is time dependent), and
the system slowly evolves to a desired
final state \cite{adia1}. Several applications of this have been
considered \cite{appln}.
Using this framework, it was shown that
Grover's search algorithm can be efficiently implemented \cite{rc}.
In this paper, we show that Deutsch's algorithm can be implemented
as well, by choosing a suitable initial state and a Hamiltonian
which evolves that state. Then a single measurement of the final state suffices
to determine whether the function $f$ is constant or balanced.
Finally, we show that the results can be extended to the
Deutsch-Jozsa algorithm involving $n$-qubits.

Let us begin with a  $2$-level system, e.g. a spin $1/2$ particle, 
with the basis kets
$\{ |0\ra, |1\ra \}$.  We define the `initial' and `final' Hamiltonians
$H_0$, $H_1$ respectively as:
\ba
H_0 &=& I - |\psi_0\ra \langle \psi_0 |      \la{h0} \\
H_1 &=& I - |\psi_1\ra \langle \psi_1 |   \la{h1}   \\
\ea
where the initial and final state vectors are given respectively by:
\ba
| \psi_0 \ra &=& \f{1}{\sqrt{2}} \le( |0 \ra + | 1  \ra  \ri)
\la{psi0} \\
| \psi_1 \ra &=& \a |0 \ra + \b | 1  \ra
\ea
with
\ba
\a &=& \f{1}{2}  \le| (-1)^{f(0)} + (-1)^{f(1)} \ri|  \\
\b &=& \f{1}{2}  \le| (-1)^{f(0)} - (-1)^{f(1)}  \ri|~~.
\ea
Note that  the Hamiltonians in the above are implicitly given in terms of 
some fundamental energy scale, ${\bar E}$, say, whose value is 
determined by the physical system used to construct the states. 
This energy scale has a natural time scale associated with it, 
namely ${\bar T} =\hbar/{\bar E}$,
which will play an important role later. The following relations will also
prove useful:
\ba
\a + \b &=& 1 \la{a1} \\
\a^2 = \a~~,~~\b^2 &=& \b \la{a3}  \\
\&~~~~~~  \a \b &=& 0~~. \la{a2}
\ea
Thus, when $f$ is a constant, $\a=1, \b=0$, and vice-versa.
We assume that the system is initially in a state $|\psi_0\ra$
and is evolved by the following time dependent Hamiltonian:
\be
 H(t) = (1-s(t))~H_0 + s(t)~H_1
\la{ht}
\ee
In general $s(t)$ is an arbitrary function of the time $t$, such that 
$s(0)=0$ and $s(1)=1$. Thus, $ H(0) = H_0$ and
${H}(1) = H_1$. For the present
we assume that $s(t)$ is linear in $t$, namely $s=t/T$,
where $T$ is the total time for which
the system is evolved. 

It follows from the adiabatic theorem that at $t=T$, the system would
be in the state $|\psi_1 \ra$,  with very high probability
$(1-\e^2)^2$ (where $\e$ is a small number),
provided the evolution is adiabatic \cite{mess,adia},
i.e.
\be
\f{| \langle \f{dH}{dt} \ra |}{g_{\min}^2}  \leq \e
\la{adia}
\ee
where the time $t$ is implicitly given in units of ${\bar T}$.
The lower bound on the evolution time $T$ is implicitly given by the condition 
(\ref{adia}) as will be seen later and
\ba
\le \langle \f{dH}{ds}  \ri \ra &:=&
\le \langle E_+,s \le | \f{dH}{dt} \ri | E_-,s  \ri \ra \\
\& ~~~\le \langle \f{dH}{dt} \ri \ra
&=& \f{ds}{dt} \le \langle \f{d{ H}}{ds} \ri \ra
\ea
$E_+(s)$ and $E_-(s)$ being the two time dependent eigenvalues
of ${ H} (s)$, with corresponding eigenvectors
$|E_+,s\ra$ and $|E_-,s\ra$ respectively, and
$$ g_{\min} :=  \min_{0 \leq s \leq 1 } [ E_+(s) - E_-(s) ] $$
Of course, the quantities $<dH/dt>$ and $g_{\min}$ should be
non-vanishing. We will show that this is indeed the case.

Thus, under adiabatic quantum evolution, the initial wave function
of the system will evolve to either $|0\ra$ or $|1\ra$, almost
with certainty, and by making a {\it single} measurement of
the state at the end, one can determine unambiguously whether
the nature of the function $f$ is. Namely, if the outcome of
measurement is $|0\ra$, then $f$ is constant and if it is
$|1\ra$, then it is balanced.

The matrix elements of ${ H}$ in the $\{|0\ra, |1\ra \}$
basis are:
\begin{displaymath}
{ H} (s) =
\left( \begin{array}{cc}
1/2 + s(\b -1/2) & ~~~~~~~-1/2(1-s) \\
{} & {} \\
-1/2(1-s) & ~~~~~~~1/2 + s(\a-1/2)
\end{array} \right)
\end{displaymath}
The corresponding eigenvalues are:
\be
E_{\pm} (s) = \f{1}{2} \le[1 \pm \sqrt{1-2s + 2s^2 }  \ri]
\ee
Note that the eigenvalues are independent of $\a$ and $\b$,
and are identical to those found in \cite{rc}. This is
a consequence of the relations (\ref{a1}) - (\ref{a2}).
It follows that
$$ \D E(s) = E_+(s) - E_-(s) = \sqrt{1-2s+2s^2} ~~,$$
which is non-zero for all values of $s$ and
\be
 g_{\min} = \D E(s=1/2) = \f{1}{\sqrt{2}}
\la{gmin2}
\ee
In addition, the two orthonormal eigenvectors are:
\begin{displaymath}
|E_\pm, s \ra  =
k_{\pm} \left( \begin{array}{c}
1-s \\
{}  \\
{} \\
 \le [ (1-2\a)s \mp \sqrt{1-2s+2s^2 }  \ri]
\end{array} \right)
\end{displaymath}
where
$$ k_{\pm} =
 {2}^{-1/2}
 \le[  [(1-2s+2s^2) \pm  (2\a -1) s \sqrt{1-2s + 2s^2}  \ri]^{-1/2}
$$
with which one can get:
\be
\le| \le \langle \f{dH}{dt} \ri \ra \ri|
= \f{1}{T}~~ \f{1}{2 \sqrt{1-2s + 2s^2} }~~.
\la{dhdt}
\ee
Note that this too is independent of $\a$ and $\b$ although
$|E_\pm,s \ra$ is not. Also, note that for any value of $s$,
the above quantity is non-vanishing and of order $1/T$.
At the final time $t=T$, the eigenstates are
$|0\ra$ and $|1\ra$ respectively, and depending on the
value of $\a, \b$, the system evolves to one of them.

Substituting (\ref{gmin2}) and (\ref{dhdt}) is (\ref{adia}), we get the
following relation:
\be
T \geq \f{1}{\e}
\la{T}
\ee
which gives an estimate of the time for which the initial
state $|\psi_0\ra$ must be evolved via the adiabatic Hamiltonian
(\ref{ht}) to attain an accuracy of order $\e$ of the final result.
For example, if we want the final state to be the $|\psi_1 \ra $ with
accuracy of  $90 \%$, then the minimum evolution time should be order of
$T \approx 1/\sqrt{1- \sqrt{0.9} } \approx 4.4 $, in units of ${\bar T}$.  

A few comments are in order here. Instead of the starting with
the initial state (\ref{psi0}), one can in general start with an
arbitrary initial state of the form
\be
|\psi_0 \ra = a |0\ra + b|1\ra~~,
\la{psi0'}
\ee
with $|a|^2 + |b|^2 =1 $, 
and evolve the system with the Hamiltonian (\ref{ht}). The end result
is expected to remain unchanged, since the ground state of
the final Hamiltonian is still $|\psi_1\ra$, to which the system
will eventually tend.
If one starts with the state (\ref{psi0'}), then $\D(s)$ and
$\langle dH/dt \ra $ are respectively:
\ba
\D (s) &=& \sqrt{1 - 4 (a^2 \b + b^2 \a ) s(1-s) }
\la{gmin1}   \\
\le | \le \langle \f{dH}{dt}  \ri \ra \ri |
&=& \f{1}{T}~~ \f{ab}
{   \sqrt{1- 4 (a^2\b + b^2\a) s(1-s) }   }
\la{dhdt1}
\ea
Substituting (\ref{gmin1}), (\ref{dhdt1}), $s=1/2$ in (\ref{adia})
and simplifying we get:
%
%
\be
T \geq \f{1}{\e} \f{ab}{ \sqrt{b^2(1-a^2)  + \a (a^2 - b^2) } }~~.
\ee
Thus depending on whether $\a=0$ of $\a=1$ (although this value is
a priori unknown), we get respectively:
\be
T \geq \f{1}{\e}~\f{a}{b}~~~~\mbox{or}~~~~
T \geq \f{1}{\e}~\f{b}{a}
\ee
Thus for extreme asymmetric values of $a$ and $b$ (e.g. $a \approx 0$
and $b \approx 1$), the evolution into the final state would either
take place in a very short or a very long time. But as the value
of $\a$ is not known, one would have wait for the greater of two values
before making the measurement. Equivalently, for very small $a$ or $b$,
$g_{\min}$ becomes very small for some value of $\a$, which is
contrary to what the validity of adiabatic theorem requires.
Consequently, the `optimal' values
for which the evolution time is independent of $\a, \b$ is given by
$$ a = b = \f{1}{ \sqrt{2} } $$
which is what we started with.

The above procedure can be generalized to Boolean functions of the
form :
$$ f : \{0,1\}^n \rightarrow \{0,1 \} $$
by making use of $n$ qubits instead of a single one \cite{d2,review1}.
In accordance with the Deutsch-Jozsa problem,
we assume that it is `promised' that the function is either
constant (i.e. all outputs are identical) or balanced (i.e. has an equal number
of $0$s and $1$s as outputs), and the task is to find which of the
above it actually is \cite{d2,review1}.
The basis states now are $\{|0\ra, |1\ra, \dots, |N-1\ra \}$
(with $N=2^n$). Now we choose the normalized initial and final states to be:
\be
|\psi_0 \ra = \f{1}{\sqrt{N}} \sum_{i=0}^{N-1} |i\ra
\ee
\be
|\psi_1 \ra  = \a |0\ra + \f{\b}{\sqrt{N-1}} \sum_{k=1}^{N-1} |k\ra
\ee
with
\ba
\a &=& \f{1}{N} \le|    \sum_{x \in \{ 0,1 \}^n} (-1)^{f(x)}   \ri|   \\
\b &=& 1- \a
\ea
Once again, if $f(x)$ is constant then $\a=1$ and $\b=0$ and vice-versa.
Thus after the required running time, if a measurement of the final 
state yields $|0\ra$, then $f(x)$ is constant and if it does not 
yield $|0\ra$, then it is balanced. 
The properties (\ref{a1} - \ref{a2}) continue to be valid.
Also, $H_0, H_1$ and ${ H}(s)$ are still given by (\ref{h0}),  (\ref{h1}) and
(\ref{ht}) respectively.

In the chosen basis,
the adiabatic Hamiltonian ${H}(s)$ is now given
by the following $N \times N$ matrix:
\begin{displaymath}
{H} (s) =
\left( \begin{array}{ccccc}
1 - \f{1-s}{N} -\a s & ~~~~~~ -\f{1-s}{N} ~~~~~ & -\f{1-s}{N}&  ~~~~~~\dots ~~~~&  -\f{1-s}{N} \\
{} & {} \\
-\f{1-s}{N} & ~~~~~~~1 - \f{1-s}{N} - \f{s\b}{N-1}  & -\f{1-s}{N} - \f{s\b}{N-1}  & ~~~~~~\dots ~~~~ & -\f{1-s}{N} - \f{s\b}{N-1}   \\
{} & {} & {} &{} & {} \\
-\f{1-s}{N}  & ~~~~~~ -\f{1-s}{N} - \f{s\b}{N-1}~~~~& 1 - \f{1-s}{N} - \f{s\b}{N-1} ~~~ & ~~~~~\dots
~~~& -\f{1-s}{N} - \f{s\b}{N-1} \\
{} & {} & {} & {}  & {} \\
~~~\dots ~~~~ & ~~~~\dots~~~~~ & ~~~~\dots~~~~~ &~~~~~~~\dots~~~~~&~~~~\dots~~~~ \\
{} & {} \\
 -\f{1-s}{N} & -\f{1-s}{N} - \f{s\b}{N-1} & ~~~~-\f{1-s}{N} - \f{s\b}{N-1} ~~~~&~~~~~~\dots~~~~ & 1 - \f{1-s}{N} - \f{s\b}{N-1}
\end{array} \right)
\end{displaymath}
It can be shown that the highest eigenvalue of the above Hamiltonian
is $1$, which is $(N-2)$-fold degenerate,
and the two remaining distinct eigenvalues (both less than $1$)
are:
\be
E_{\pm} (s) =
\f{1}{2} \le[1 \pm  \sqrt{ 1- \f{4s(1-s)}{N} \le( \b + \a(N-1)  \ri) }  \ri]~~.
\ee
Thus, 
\be
\D E(s) = E_+(s) - E_-(s) =
\sqrt{ 1- \f{4s(1-s)}{N} \le( \b + \a(N-1)  \ri) } \neq 0~~~,
\la{deltaE}
\ee
implying
\be
 g_{\min} = \D E(s=1/2) =
\sqrt{ 1 - \f{1}{N}  \le( \b + \a(N-1)  \ri) }
\la{gmin}
\ee
In addition, it can be shown that:
\be
\le| \le \langle \f{d{ H}}{ds} \ri \ra \ri|
= \f{\sqrt{N-1}}{N}~
\f{1}{ \sqrt{ 1-\f{4s(1-s)}{N} [\b + \a(N-1)]  }  }~~~.
\la{dhdt2}
\ee
Thus for $s=t/T$, from condition (\ref{adia}) it follows that (for large $N$)
\be
T \geq \f{N}{\e}
\la{t1}
\ee
which shows that the evolution time scales as $N$, the number of qubits.

However, following \cite{rc} if we assume an evolution
with a general $s(t)$ we obtain the adiabaticity condition 
(\ref{adia}) that must be satisfied at any given instant of time $t$:
\be
\f{ds}{dt} \leq \e \f{ \le[ E_+(s) -  E_-(s) \ri]^2}
{\le| \le\langle  \f{dH}{ds} \ri\ra \ri|} 
\la{sat1}
\ee
Substituting (\ref{deltaE}) and (\ref{dhdt2}), we get 
(here $\c \equiv 4(\b + \a (N-1) ) /N$) :
\be
\int_0^T dt \geq \f{1}{\e}~\f{\sqrt{N-1}}{N} 
\int_o^1 \f{ds}{[1- s\c + s^2 \c ]^{3/2}}  
\ee
The expression on the r.h.s. can be integrated, and is equal to
$$
 \f{1}{\e} \le[ \f{\sqrt{N-1}}{N}~\f{1}{\c^{3/2}}~
\f{2s-1}{(\c^{-1} - 1/4) \sqrt{s^2 - s + \c^{-1}}  } 
\ri]
$$
Inserting the limits of $s$, we get the following bound:
\be
T \geq \f{1}{\e}~\f{\sqrt{N-1}}{N} \f{1}{\c (1 - \c/4)}~~,
\ee
It can be verified that for $N >> 1 $, this lower bound is 
\be
T \geq \f{\sqrt{N}}{\e} 
\la{bound1}
\ee
which is a quadratic improvement over the previous bound (\ref{t1}). 
Also since relation (\ref{sat1}) has to be satisfied
at every instant, the bound (\ref{bound1}) is optimal. 

To conclude,
in this paper, we have implemented Deutsch's algorithm using
adiabatic quantum evolution by a Hamiltonian which takes a given
initial state to a final state and such that the final state
depends on the nature of
the function $f$. In particular, if the function is constant, the
final state is $|0\ra$ with a very high probability, and if it is
balanced then the outcome is $|1\ra$ almost with certainty.
Then a measurement on the
final state helps to determine the nature of $f$. We have also
estimated the required evolution time for a given accuracy of
the result. Finally, we have generalized the result for the
Deutsch-Jozsa problem, using $n$-qubits, and found that the 
number of time steps required to solve the problem 
scales as $\sqrt{N}$, where $N = 2^n$. Although this is a marginal improvement over the classically required exponential time (of order $N$), it does not match the polynomial time that is achievable using standard quantum computational techniques\cite{d2}.
It would be
interesting to compare adiabatic and standard quantum computational methods for other algorithms to see whether this difference in computational time is the exception rather than the rule. We hope to report on this elsewhere.

\vs{.4cm}
\no
{\bf Acknowledgements}

\no
We would like to thank J. Currie, R. Laflamme and 
V. Linek for useful discussions. 
This work was supported in part by the Natural Sciences and 
Engineering Research Council of Canada.

\end{document}